\documentclass[conference]{IEEEtran}
\IEEEoverridecommandlockouts
\usepackage{cite}
\usepackage{amsmath,amssymb,amsfonts}
\usepackage{graphicx}
\usepackage{textcomp}
\usepackage{xcolor}
\usepackage{latexsym}
\usepackage{float}
\usepackage{caption}
\usepackage{subcaption}
\usepackage{algorithm}
\usepackage{comment}
\usepackage[noend]{algpseudocode} 
\usepackage{multirow}
\usepackage{url}

\def\BibTeX{{\rm B\kern-.05em{\sc i\kern-.025em b}\kern-.08em
    T\kern-.1667em\lower.7ex\hbox{E}\kern-.125emX}}

\begin{document}

\title{
Membership Inference Attacks against \\ Diffusion Models
}

\author{\IEEEauthorblockN{Tomoya Matsumoto}
\IEEEauthorblockA{\textit{Osaka University} \\ Osaka, Japan \\ t-matsumoto@ist.osaka-u.ac.jp}
\and
\IEEEauthorblockN{Takayuki Miura}
\textit{Osaka University}, \\
\IEEEauthorblockA{\textit{NTT Social Informatics Laboratories} \\ Tokyo, Japan \\ takayuki.miura.br@hco.ntt.co.jp}
\and
\IEEEauthorblockN{Naoto Yanai}
\IEEEauthorblockA{\textit{Osaka University} \\ Osaka, Japan \\
yanai@ist.osaka-u.ac.jp}
}

\maketitle


\begin{abstract}
Diffusion models have attracted attention in recent years as innovative generative models. 
In this paper, we investigate whether a diffusion model is resistant to a membership inference attack, which evaluates the privacy leakage of a machine learning model. 
We primarily discuss the diffusion model from the standpoints of comparison with a generative adversarial network (GAN) as conventional models and hyperparameters unique to the diffusion model, i.e., timesteps, sampling steps, and sampling variances. 
We conduct extensive experiments with DDIM as a diffusion model and DCGAN as a GAN on the CelebA and CIFAR-10 datasets in both white-box and black-box settings and then show that the diffusion model is comparably resistant to a membership inference attack as GAN.
Next, we demonstrate that the impact of timesteps is significant and intermediate steps in a noise schedule are the most vulnerable to the attack. 
We also found two key insights through further analysis. 
First, we identify that DDIM is vulnerable to the attack for small sample sizes instead of achieving a lower FID. 
Second, sampling steps in hyperparameters are important for resistance to the attack, whereas the impact of sampling variances is quite limited. 
\end{abstract}

\begin{IEEEkeywords}
diffusion model, membership inference attack, GAN, hyperparameter, privacy
\end{IEEEkeywords}

\section{Introduction} 

\subsection{Motivation} \label{sec:motivation}

In machine learning, generative models have been actively studied. 
As a breakthrough in recent years, diffusion models~\cite{Nonequilibrium,DDPM} were discovered as new generative models. 
Diffusion models can generate more plausible images and texts than existing generative models such as a generative adversarial network (GAN). 
Remarkably, diffusion models have outperformed GANs on academic benchmarks in the past years~\cite{dhariwal2021diffusion} and are expected to replace GANs in a variety of applications~\cite{ramesh2022hierarchical,rombach2022high,saharia2022photorealistic,nichol2022glide}. 

We emphasize that discussion on privacy violations for training data of diffusion models~\cite{hu2023membership,carlini2023extracting,duan2023are} is a very recent issue and hence is still insufficient.
In general, training generative models requires large amounts of training data, and, for example, models that generate disease histories as sensitive information also attract attention~\cite{choi17ageneratingmulti-label,YI2019generative}. 
Understanding privacy violations under realistic threats for generative models is crucial for using generative models in the real world~\cite{Hilprecht2019montecarlo}.
Consequently, privacy violations of diffusion models should be discussed in more detail.

In this paper, we discuss a membership inference attack~\cite{ShokriSSS17} against diffusion models, which is an important criterion for evaluating privacy violations. 
Informally, a membership inference attack aims to identify whether a data record was used to train a machine learning model. 
This attack is discussed for evaluating privacy violations in the past years~\cite{conti2022label-onlymembership,li2022auditingmembershipinference,ye2022enchancedmembership}, and hence it is important to discuss the attack for diffusion models in order to understand potential threats of the model. 
We then try to answer the following question: \textit{How vulnerable are diffusion models to a membership inference attack?}

We note that this question is \textit{non-trivial} because the data generation logic by diffusion models differs from conventional models such as GANs. 
According to earlier literature~\cite{Hayes2019logan,Hilprecht2019montecarlo,mukherjee2021privgan,stadler2021groundhog}, the architecture of a generative model affects the attack success rate of a membership inference attack.
For example, for an attack against GAN, an attacker can utilize a discriminator model that is a counterpart to the generative model~\cite{mukherjee2021privgan}.
By contrast, diffusion models do not contain such a model. 
Meanwhile, diffusion models have unique hyperparameters, i.e., timesteps, sampling steps, and sampling variances, that have never been contained in conventional models. 
It indicates that we no longer know how the membership inference attack on diffusion models varies compared to the traditional models.
These standpoints were not addressed in the existing works~\cite{hu2023membership,duan2023are,carlini2023extracting}. 
(See Section~\ref{sec:related_AttackOnDiffusion} for more detail.)




\subsection{Contributions}

In this paper, we shed light on the impact of a membership inference attack on diffusion models through extensive experiments. 
Our primary discussions are on comparison with a GAN as a conventional generative model and hyperparameters of the diffusion model, i.e., timesteps, sampling steps, and sampling variances. 
In particular, through extensive experiments with denoising diffusion implicit models (DDIM) as a diffusion model and deep convolutional GAN (DCGAN) as a GAN on the CelebA and CIFAR-10 datasets in both white-box and black-box settings, we show that the diffusion model is comparably resistant to a membership inference attack as GAN. 
We then identify the impact of timesteps on the attack. 
Intermediate steps in a noise schedule are the most vulnerable to the attack rather than the beginning or the final step. 
(See Section~\ref{sec:experiments} for detail.)

We also provide two key insights into a membership inference attack on the diffusion model through further analysis for overfitting compared to GAN and hyperparameters for sampling, i.e., sampling steps and sampling variances.
First, for overfitting, we show that DDIM is vulnerable to the attack if the number of training samples is small since DDIM is well-trained compared to DCGAN. 
Second, sampling steps in hyperparameters are quite important for resistance to the attack in contrast to sampling variances, which are irrelevant to the attack. 
(See Section~\ref{sec:discussion} for detail.) 
Our source code is publicly available (\url{https://github.com/fseclab-osaka/mia-diffusion}). 

\section{Related Work}

This section describes related works of membership inference attacks, including GANs and diffusion models. 






\subsection{Membership Inference} 
\label{sec:background_MIA}

A membership inference attack is a kind of attack whereby an adversary infers whether a particular example was contained in the training dataset of a model~\cite{shokri2017membership,ML-Leaks2019ndss,carlini2022firstprinciples}. 
A model vulnerable to the attack potentially contains threats to privacy leakage, and hence recent works discuss membership inference attacks for various machine learning models~\cite{Hayes2019logan,conti2022label-onlymembership,li2022auditingmembershipinference}. 
There are two settings~\cite{ShokriSSS17}, i.e., the white-box setting where an adversary has access to model parameters, and the black-box setting where he/she utilizes only outputs of the model. 

A typical approach for membership inference attacks is to leverage the large divergence between the loss distribution over members and non-members~\cite{song2019membershipinference}. 
The divergence can be embedded by an adversary. 
For instance, privacy leakage, including membership inference attacks, can be more effective by training a model with poisoning samples~\cite{tramer2022truth,hidano2018modelinversion,wang2022poisoning-assisted,mahloujifar2022property}. 

Membership inference attacks can be prevented by differential privacy~\cite{Dwork06} where gradients are perturbed~\cite{abadi2016deep,yeom2018privacy}. 
Since differential privacy often deteriorates inference accuracy, several works evaluated differential privacy on membership inference attacks in a quantitative fashion~\cite{jayaraman2019evaluating,jagielski2020auditing}. 


\subsection{Attacks on GANs} 
\label{sec:relatedwork}

    Related works are membership inference attacks against GANs~\cite{Hayes2019logan,mukherjee2021privgan,Hilprecht2019montecarlo,GAN-Leaks}. 
    LOGAN was presented as the first attack against GANs in both white-box and black-box settings. 
    LOGAN in the white-box setting utilizes outputs from the discriminator of a target model. 
    Concurrently, the Monte Carlo attack~\cite{Hilprecht2019montecarlo} was presented as an attack independent of a target model's architecture in the black-box setting. 
    The above attacks are also utilized as oracles in the subsequent work~\cite{mukherjee2021privgan}. 


    GAN-Leaks~\cite{GAN-Leaks} is the state-of-the-art attack against GANs in the black-box setting, to the best of our knowledge. 
    It no longer requires the setting of parameters compared to the above attacks and hence can be used in diffusion models. 
    GAN-Leaks also showed a new attack whereby an adversary has access to the internal of a target model's generator. 

    We utilize LOGAN in the white-box setting and GAN-Leaks in the black-box setting to compare a diffusion model with a GAN, respectively. 
    As described above, LOGAN is the only work on GANs in the white-box setting while GAN-Leaks can be used in diffusion models. 
    We note that there is no work that is applicable to diffusion models in the white-box setting.


\subsection{Attacks on Diffusion Models}\label{sec:related_AttackOnDiffusion}

    We note that there is an attack by Wu et al.~\cite{wu2022membershipagainsttext-to-image} against diffusion models. 
    Their attack focuses on text-to-image generation models. 
    In contrast, we focus on typical image generation models. 
    Namely, the underlying problem is quite different. 
    It is considered that the existing attacks against GANs are rather close to our work than Wu et al.~\cite{wu2022membershipagainsttext-to-image}.  

Concurrently, there are a few works~\cite{hu2023membership,duan2023are,carlini2023extracting} on typical image generation models based on diffusion models, which are the closest to ours. 
As the main difference from these works, we discuss hyperparameters of diffusion models in detail, which are our main results. 
In particular, Hu et al.~\cite{hu2023membership} and Duan et al.~\cite{duan2023are} did not discuss comparison with GANs. 
Carlini et al.~\cite{carlini2023extracting} compared with GANs, but did not discuss hyperparameters, i.e., sampling steps and sampling variances. 
We primarily focus on comparison with GANs and discussion on the hyperparameters. 
Our work is thus concurrent with the above works. 
Interested readers are also encouraged to read the existing works~\cite{hu2023membership,duan2023are,carlini2023extracting}.


\section{Membership Inference Attacks against Diffusion Model} 
\label{sec:MIAtoDiffusion}

This section presents our membership inference attacks against a diffusion model. 
To this end, we first formalize the adversary setting of the attacks and the detail of a diffusion model. 
We then describe our investigation method and the key question of this paper in detail. 

\begin{figure}[t!]
   \centering
    \includegraphics[scale = 0.31]{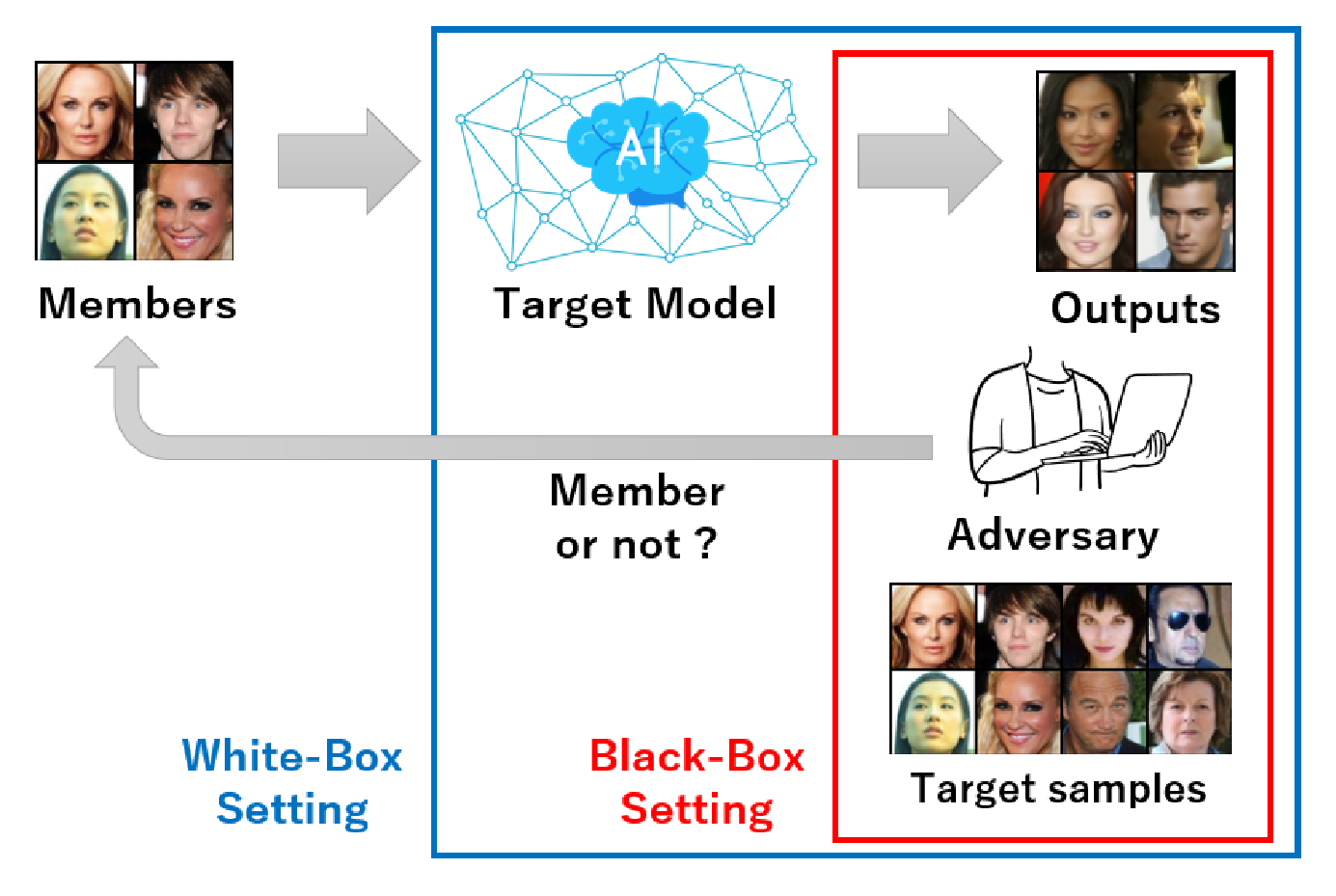}
    \caption{Overview of our membership inference attacks.}
    \label{fig:ovewview}
\end{figure}



\subsection{Formalization}
\label{sec:adversary_setting}

A membership inference attack in this paper is defined as a game between an adversary $\mathcal{A}$ and a challenger $\mathcal{C}$. 
We note that the adversary's knowledge about the target model is different for white-box and black-box settings. 
White-box setting means that an adversary has access to an architecture of the model and its parameters while black-box setting means that an adversary cannot obtain these information. 

We first denote by $\mathbb{R}$ a set of real numbers, 
by $\mathcal{D}$ a set of data samples, and by $\mathcal{M}$ a set of generative models. 
Then, a generative model $M \in \mathcal{M}$ is defined as a function $M: \mathbb{R}^r \rightarrow \mathcal{D}$, where $r$ is an arbitary integer. 
The game is as follows:
\begin{enumerate}
\item $\mathcal{C}$ chooses a dataset $D\subseteq \mathcal{D}$. 
    \item $\mathcal{C}$ chooses a bit $b\leftarrow \{0,1\}$. 
    If $b=1$, $\mathcal{C}$ chooses a sample $x\in D$; otherwise, $\mathcal{C}$ chooses $x\in \mathcal{D}\backslash D$ other than $D$. 
    $\mathcal{C}$ sends $x$ to $\mathcal{A}$. 
    \item  $\mathcal{C}$ trains $M$ with the dataset $D$. 
    \item $\mathcal{A}$ obtains  $x_i = M(z_i)$  for any $z_i \in \mathbb{R}^r$. 
    \textcolor{red}{Furthermore, $\mathcal{A}$ also receives $M$ and its auxiliary information\footnote{In the case of GANs, a discriminator is the auxiliary information.}.}
    \item $\mathcal{A}$ returns a bit $b'\in \{0,1 \}$. 
    If $b=b'$, $\mathcal{A}$ wins the game. 
\end{enumerate}

The difference between the white-box and black-box settings is represented as the sentences with red color, and only the white-box setting contains this sentence. 
As described in Section~\ref{sec:background_MIA}, an adversary in the white-box setting has access to the internal of a target model $M$: 
in this paper, he/she can have access to internal networks, which are utilized in the training of $M$ and unnecessary for data generation, such as discriminators in GANs as auxiliary information. 
On the other hand, an adversary in the black-box setting receives only the generated samples $x_i$ from a target model $M$. 



We adopt area-under-the-ROC-curve (AUCROC)~\cite{song2019auditing},  attack success rate (ASR)~\cite{ShokriSSS17}, and true positive rates (TPR) at low false positive rates (FPR)~\cite{carlini2022firstprinciples}, used as TPR at 1\%FPR in this paper, as evaluation metrics of the attacks. 
We also adopt the Frechet inception distance (FID)~\cite{TTUR_fid} to evaluate the quality of generated images. 
A lower score for FID means that generated images are of a higher quality. 


\subsection{Detail of Diffusion Model} 
\label{sec:DetailofDiffusion}

Membership inference attacks in this paper depend on loss values of a diffusion model. 
We describe the detail of the loss function, including the diffusion process and denoising one. 

We describe the denoising diffusion probabilistic models (DDPM)~\cite{DDPM} below for simplicity because DDIM has nearly the same process.
The diffusion process is described as:
\begin{eqnarray}
\label{eq:forward1}
q(x_{t}|x_{t-1})=\mathcal{N}(x_{t};\sqrt{1-\beta_{t}}x_{t-1},\beta_{t}I).
\end{eqnarray}
According to the noise schedule $\beta$, the observed data $x_{0}$ are transformed into noise $x_{T}$ by adding Gaussian noises over $T$ steps.
The denoising process is described as:
\begin{eqnarray}
\label{eq:reverse2}
p_\theta(x_{t-1} \vert x_t) = \mathcal{N}(x_{t-1}; \mu_\theta(x_t, t), \Sigma_\theta(x_t, t)). 
\end{eqnarray}
The data are reconstructed by repeatedly removing noises from noise $x_{T}$ drawn from a multivariate standard Gaussian distribution. 

The model is trained to maximize the variational lower bound of the log marginal likelihood $\log p_\theta(x_0)$.
Therefore, the loss function $L$ is expressed as:
\begin{eqnarray}
\label{eq:loss}
L := \mathbb{E}_q \Big[ \log \frac{q(x_{1:T}\vert x_0)}{p_\theta(x_{0:T})} \Big] \geq \mathbb{E} \Big[ - \log p_\theta(x_0) \Big]. 
\end{eqnarray}
In practice, we use the simplified loss function shown as:
\begin{eqnarray}
\label{eq:loss_simple}
L_{simple} := \mathbb{E}_{t, x_{0}, \epsilon} \Big[ \| \epsilon - \epsilon_\theta (\sqrt{\bar{\alpha}_{t}}x_{0}+\sqrt{1-\bar{\alpha}_{t}}\epsilon, t) \| ^2 \Big],
\end{eqnarray}
which represents the noise estimation error of the data $x_{t}$, using the notation $\bar{\alpha}_{t}:=\prod_{s=1}^{t}(1-\beta_{s})$.

In training, $\theta$ is optimized for the above loss function. 
It is also utilized for the membership inference attack because  small loss values mean that the target sample is in-member. 

Meanwhile, the sampling of DDIM is represented as:
\begin{eqnarray}
\label{eq:sampling}
x_{t-1} = \sqrt{\bar{\alpha}_{t-1}}\left(\frac{x_t-\sqrt{1-\bar{\alpha}_t}\epsilon_\theta(x_t,t)}{\sqrt{\bar{\alpha}_t}}\right) \nonumber \\
+\sqrt{1-\bar{\alpha}_{t-1}-{\sigma_t}^2}\cdot\epsilon_\theta(x_t,t)+\sigma_t\epsilon_t,
\end{eqnarray}
where $\sigma$ controls the randomness of the sampling. 
We can generate images for any subset of timesteps $\tau \subset \{ 1,...,T \}$ and $\sigma$ is represented with a hyperparameter $\eta$ as: 
\begin{eqnarray}
\label{eq:eta}
\sigma_{\tau_i}(\eta) := \eta\sqrt{(1-\bar{\alpha}_{\tau_{i-1}})/(1-\bar{\alpha}_{\tau_i})}\sqrt{1-\bar{\alpha}_{\tau_i}/\bar{\alpha}_{\tau_{i-1}}}. 
\end{eqnarray}
Note that the following value is utilized for CIFAR-10 in~\cite{DDIM}:
\begin{eqnarray}
\label{eq:noisy}
\hat{\sigma}_{\tau_i} := \sqrt{1-\bar{\alpha}_{\tau_i}/\bar{\alpha}_{\tau_{i-1}}}. 
\end{eqnarray}

\subsection{Investigation Method}
\label{sec:our_method}

We describe our investigation method to evaluate membership inference attacks against a diffusion model in the white-box setting. 
As described in Section~\ref{sec:relatedwork}, there is no attack against a diffusion model in the white-box setting, whereas we utilize LOGAN~\cite{Hayes2019logan} and GAN-Leaks~\cite{GAN-Leaks}. 

The loss function of a diffusion model represents a noise estimation error as in Equation~(\ref{eq:loss_simple}). 
In general, the loss values among members are smaller than among non-members because a model learns training data to minimize them~\cite{song2019membershipinference}. 
Based on this fact, an adversary can infer whether the target samples are contained in the training dataset by computing their loss values with appropriate time $t$ (as described later).
The detail of the method is shown in Algorithm~\ref{alg:our_attack}. 

\renewcommand{\algorithmicrequire}{\textbf{Input:}}
\renewcommand{\algorithmicensure}{\textbf{Output:}}

\begin{algorithm}[t!]
    \caption{White-box attack against diffusion models}
    \label{alg:our_attack}
    \begin{algorithmic}
    \Require Target samples $x^1,...,x^{m}$, model's network $\epsilon_\theta$, time $t$, noise schedule $\bar{\alpha}_{t}:=\prod_{s=1}^{t}(1-\beta_{s})$, threshold $c$
    \For {$i = 1$ to $m$}
    \State $y^i \leftarrow 0$
    \Comment{Initialization}
    \If{$\epsilon - \epsilon_\theta(\sqrt{\bar{\alpha}_{t}}x^i+\sqrt{1-\bar{\alpha}_{t}}\epsilon,t) < c$}
    \State $y^i \leftarrow 1$ 
    \Comment{Inferring as a member}
    \EndIf
    \EndFor
    \Ensure Labels $y^1,...,y^{m}$
    \end{algorithmic}
\end{algorithm}

\subsection{Key Question}

As described in Section~\ref{sec:motivation}, we shed light on how a diffusion model is vulnerable to membership inference attacks. 
To this end, we discuss it from two standpoints: difference from generative adversarial models (GANs), and timesteps of a diffusion model. 

We first evaluate membership inference attacks, including the investigation method in the previous section, against the diffusion model and existing GANs. 
The main difference between the diffusion model and existing GANs is the number of neural networks, i.e., the diffusion model consists of a single neural network while the existing GANs consist of generators and discriminators. 
We discuss the impact of the above difference on membership inference attacks in both black-box and white-box settings.

We also discuss the impact of timesteps on membership inference attacks. 
In a diffusion model, information learned by the model is different for each timestep $t$. 
Although the performance of membership inference attacks depends on the divergence of loss distributions between members and non-members~\cite{song2019membershipinference}, the timestep $t$ when the divergence is largest is unknown.
Consequently, we evaluate membership inference attacks with respect to timestep $t$. 



\section{Experiments} 
\label{sec:experiments}

This section conducts extensive experiments on membership inference attacks against a diffusion model. 
We evaluate the attack AUC by comparing the attacks against GANs. 
We also evaluate the impact of the timesteps of a diffusion model on the attack AUC.



\subsection{Experimental Setting}
\label{sec:exp_setting}

The experimental setting in this paper is described below. 

\subsubsection{Datasets}

We utilize CIFAR-10~\cite{krizhevsky2009learning} and CelebA~\cite{celeba} datasets, which contain samples with $32\times32$ color pixels and with center-cropped $64\times64$ color pixels, respectively. 

\subsubsection{Architectures} 
We utilize the denoising diffusion implicit model (DDIM)~\cite{DDIM} as a diffusion model and the deep convolutional GAN (DCGAN)~\cite{DCGAN} as a baseline. 
The reason of the use of these architectures is that DDIM is the basis of other diffusion models and is suitable for investigating common properties of the diffusion models. 
On the other hand, DCGAN has been utilized in existing works\cite{Hayes2019logan, Hilprecht2019montecarlo, GAN-Leaks} for membership inference attacks on generative models. 

For hyperparameters, we set $T = 1000$, and the noise schedule is the cosine schedule~\cite{ImprovedDDPM} or the linear schedule~\cite{DDPM}. 
The default setting is the cosine schedule. 
Also, the sampling is performed for sampling variance $\sigma(0)$ and per 50 steps, i.e., 20 steps in total. 
The dimension of latent variables in DCGAN is 100.
A target model $M$ is trained with 12,000 samples, and the numbers of epochs are 500 for DDIM and 300 for DCGAN. 
To fairly compare the evaluation metrics for membership inference attacks, the effect of overfitting should be minimized. 
Thus, we utilize 500 epochs for DDIM and 300 epochs for DCGAN as the target model $M$.

\begin{figure}[t!]
  \begin{minipage}[b]{0.49\linewidth}
    \centering
    \includegraphics[keepaspectratio, scale=0.275]{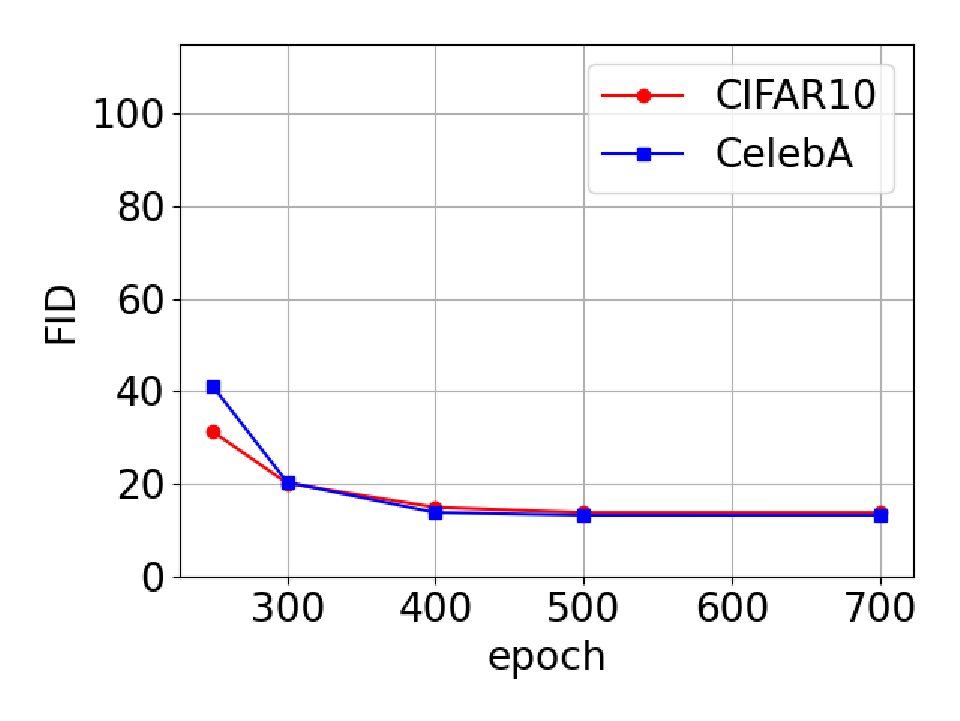}
    \subcaption{DDIM}
  \end{minipage}
  \begin{minipage}[b]{0.49\linewidth}
    \centering
    \includegraphics[keepaspectratio, scale=0.275]{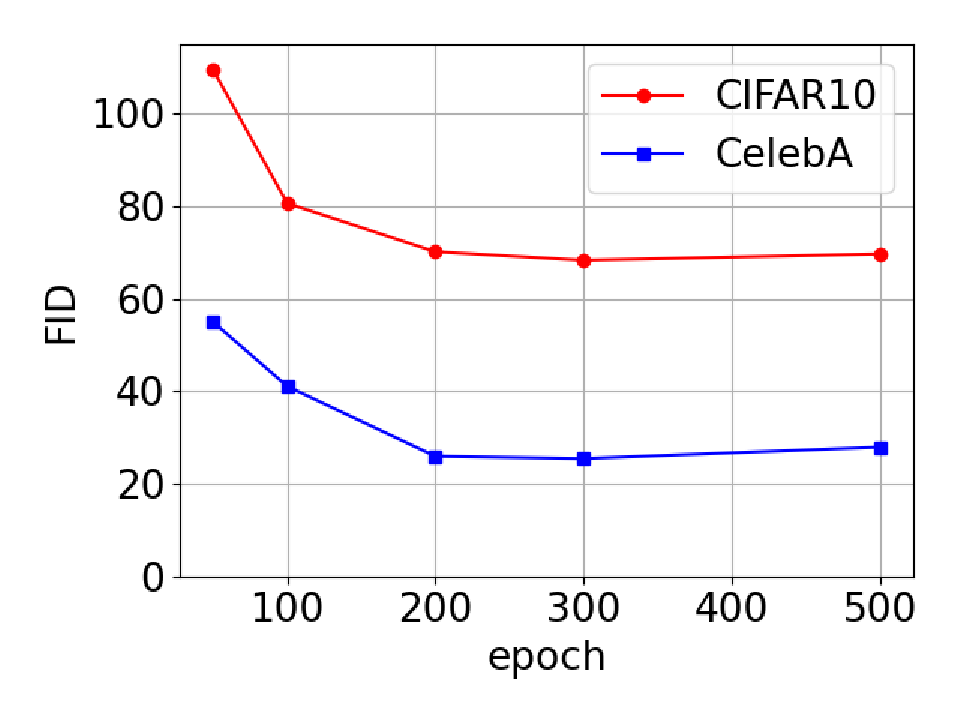}
    \subcaption{DCGAN}
  \end{minipage}
  \caption{FID scores throughout training.}
  \label{fig:fid}
\end{figure}


\subsubsection{Attack Methods}

In the white-box setting, we utilize the investigation method described in Section~\ref{sec:our_method} for the attack on DDIM, where time $t=350$, and LOGAN~\cite{Hayes2019logan} for the attack on DCGAN. 
On the other hand, in the black-box setting, we utilize the full black-box attack of GAN-Leaks~\cite{GAN-Leaks} where the number of generated images is 12,000. 



\subsection{Results}

\subsubsection{Comparison with GANs}

\begin{figure}[t!]
  \begin{minipage}[b]{0.49\linewidth}
    \centering
    \includegraphics[keepaspectratio, scale=0.275]{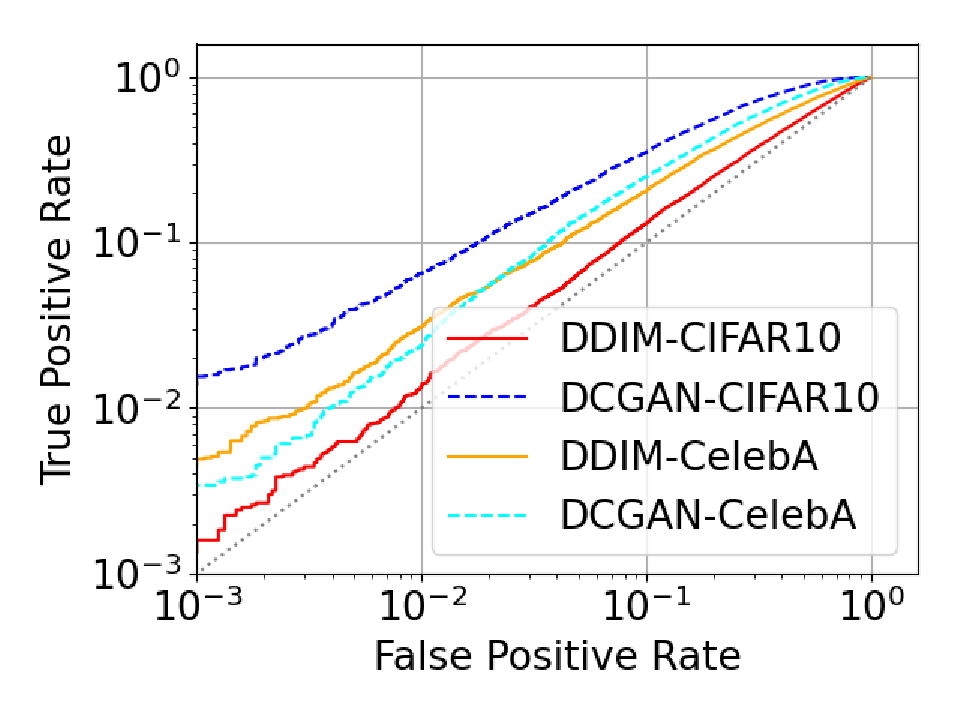}
    \subcaption{White-box setting.}
  \end{minipage}
  \begin{minipage}[b]{0.49\linewidth}
    \centering
    \includegraphics[keepaspectratio, scale=0.275]{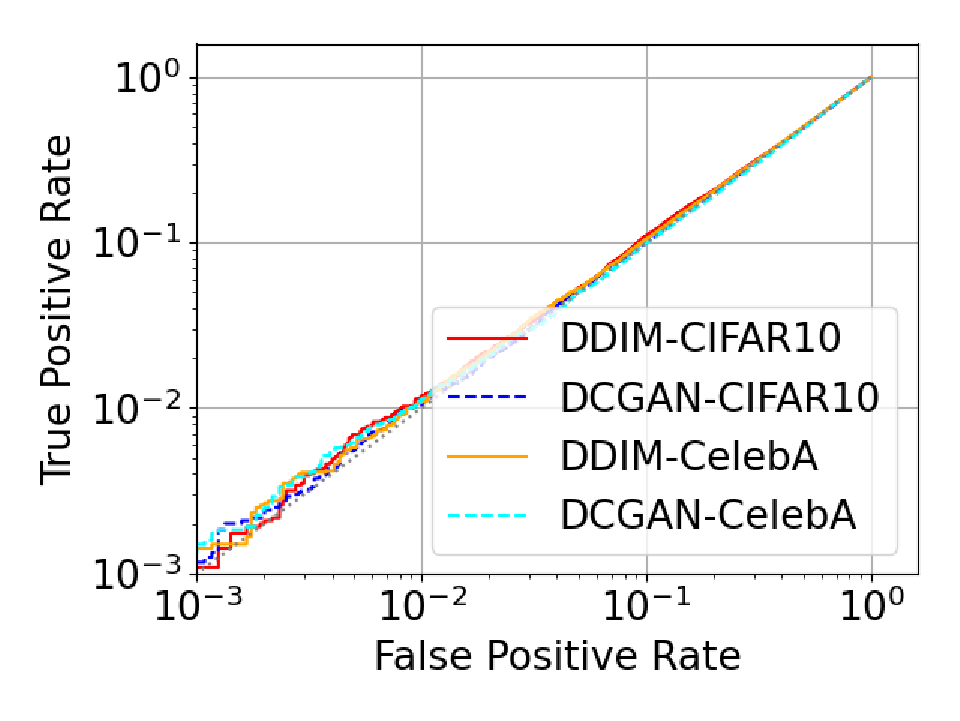}
    \subcaption{Black-box setting.}
  \end{minipage}
  \caption{ROC curves.}
  \label{fig:roc}
\end{figure}

The results of DDIM and DCGAN are shown in Fig.~\ref{fig:roc} and Tables~\ref{tab:normal}. 
According to these results, DDIM is more resistant to the membership inference attack than DCGAN on CIFAR-10 in the white-box setting. 
Furthermore, according to Table~\ref{tab:normal}, DDIM is more resistant than DCGAN in terms of AUCROC. 
Although one might think that DDIM is vulnerable according to TPR at 1\%FPR, including several low scores of FPR in Fig.~\ref{fig:roc}, on CelebA, it is considered that they are fairly comparable because each value is quite small. 
By contrast, the attack fails for all models and datasets in the black-box setting. 
As a result, it is considered that DDIM is at least resistant equivalently to DCGAN. 

\begin{table}[t!]
    \centering
    \caption{Attack performance on white-box/black-box setting.}
    \begin{tabular}{c|c|ccc}
    \hline
          \multirow{2}{*}{Dataset} & \multirow{2}{*}{Model} & \multirow{2}{*}{AUCROC} & \multirow{2}{*}{ASR} & TPR at \\ 
         & & & & 1\%FPR \\ \hline \hline
         \multirow{2}{*}{CIFAR-10} & DDIM & 0.552 / 0.503 & 0.536 / 0.499 & 1.54 / 1.19\% \\
              & DCGAN & 0.778 / 0.497 & 0.707 / 0.497 & 6.50 / 1.05\% \\ \hline
         \multirow{2}{*}{CelebA} & DDIM & 0.635 / 0.503 & 0.599 / 0.502 & 2.93 / 1.09\% \\
               & DCGAN & 0.699 / 0.495 & 0.643 / 0.494 & 2.56 / 1.13\% \\ \hline
    \end{tabular}
    \label{tab:normal}
\end{table}

\subsubsection{Attack performance for different timesteps}
\label{sec:auc_diffusion}

The investigation method in Section~\ref{sec:our_method} requires a timestep $t$ as an input to a target model $M$ in addition to samples $x^1, \ldots, x^m$. 
We show that timesteps and the noise schedule strongly affect the performance of the membership inference attacks. 

\begin{figure}[t!]
   \centering
    \includegraphics[scale = 0.33]{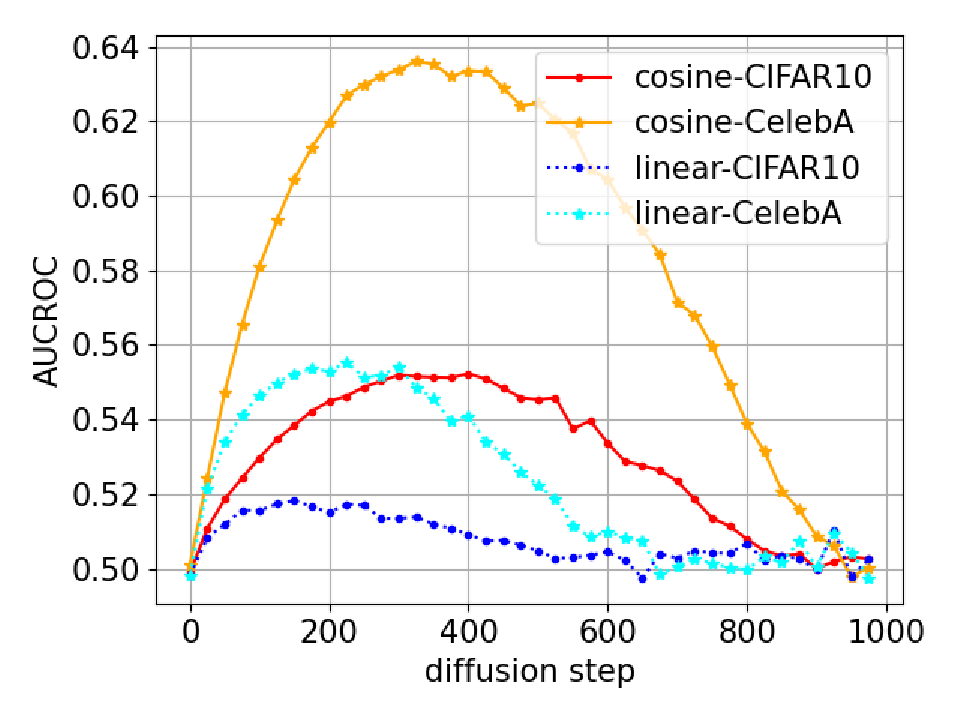}
    \caption{AUCROC at every 25 step on white-box setting.}
    \label{fig:auc}
\end{figure}

The results of AUCROC for each timestep $t$ are represented in Fig.~\ref{fig:auc}, where DDIM is trained with the cosine and linear schedules. 
According to the figure, AUCROC is maximized at $t=350$ for the cosine schedule and at $t=200$ for the linear schedule. 
We call these values \textit{best-one} AUCROC for the sake of convenience. 
For these values, the value of $\bar{\alpha}_{t}$ is 0.7. 
As might have been unexpected, these maximized values were obtained in intermediate steps. 
The best-one AUCROC for the cosine schedule is higher than the liner schedule. 
Besides, according to Table~\ref{tab:oneandall}, AUCROC with an average of loss values for all the steps, called \textit{average} AUCROC for the sake of convenience, is lower than the best-one AURCROC. 
Therefore, we use results at $t=350$ for the investigation methods in the entire experiments. 


\begin{table}[t!]
    \caption{AUCROC by best-one and average $t$.}
    \label{tab:oneandall}
    \centering
    \begin{tabular}{c|cc|cc}\hline
     & \multicolumn{2}{c|}{CIFAR-10} & \multicolumn{2}{c}{CelebA} \\
    \cline{2-5}
     & best-one & average & best-one & average \\ \hline \hline
    cosine & 0.552 & 0.535 & 0.636 & 0.597\\
    linear & 0.518 & 0.510 & 0.555 & 0.529 \\ \hline
    \end{tabular}
\end{table}

\section{Discussion}
\label{sec:discussion}

This section discusses the attack performance in terms of number of training samples and training epochs to understand the impact of overfitting on diffusion models. 
We then discuss the impacts of hyperparameters for sampling in the diffusion models.


    

    


\subsection{Impact of Overfitting} 

Overfitting strengthens membership inference attacks in general~\cite{yeom2018privacy}. 
We discuss the impact of overfitting in terms of number of training samples and epochs. 

\subsubsection{Number of training samples}

\begin{table*}[t!]
    \centering
    \caption{Attack performance for different number of training samples in white-box/black-box setting.}
    \begin{tabular}{c|r|ccr|r|ccr|r}
    \hline
         \multirow{2}{*}{Model} & Num. of  & \multicolumn{4}{c|}{CIFAR-10} & \multicolumn{4}{c}{CelebA} \\
         \cline{3-10}
          & samples & AUCROC & ASR & TPR at 1\%FPR & FID & AUCROC & ASR & TPR at 1\%FPR & FID \\ \hline \hline
              & 12,000 & 0.552 / 0.503 & 0.536 / 0.499 & 1.54 / ~1.19\% & 13.91 & 0.635 / 0.503 & 0.599 / 0.502 & 2.93 / ~1.09\% & 13.94 \\
          DDIM&  6,000 & 0.801 / 0.499 & 0.736 / 0.500 & 5.65 / ~0.90\% & 16.27 & 0.973 / 0.497 & 0.901 / 0.499 & 65.30 / ~0.90\% & 16.64 \\
              &    600 & 1.000 / 0.931 & 0.968 / 0.882 & 100.00 / 77.50\% & 28.72 & 1.000 / 0.762 & 0.995 / 0.717 & 100.00 / 47.33\% & 33.89 \\ \hline
               & 12,000 & 0.778 / 0.497 & 0.707 / 0.497 & 6.50 / ~1.05\% & 68.29 & 0.699 / 0.495 & 0.643 / 0.494 & 2.56 / ~1.13\% & 25.54 \\
         DCGAN &  6,000 & 0.983 / 0.497 & 0.899 / 0.499 & 69.80 / ~0.78\% & 79.30 & 0.776 / 0.492 & 0.694 / 0.490 & 8.33 / ~0.95\% & 31.99 \\
               &    600 & 1.000 / 0.526 & 0.988 / 0.529 & 99.50 / ~2.17\% & 189.96 & 0.940 / 0.524 & 0.873 / 0.520 & 19.83 / ~1.50\% & 227.53 \\ \hline
    \end{tabular}
    \label{tab:attack_eval_total}
\end{table*}

\begin{figure}[t!]
    \centering
    \begin{subfigure}[b]{0.49\linewidth}
        \centering
        \includegraphics[keepaspectratio, scale=0.275]{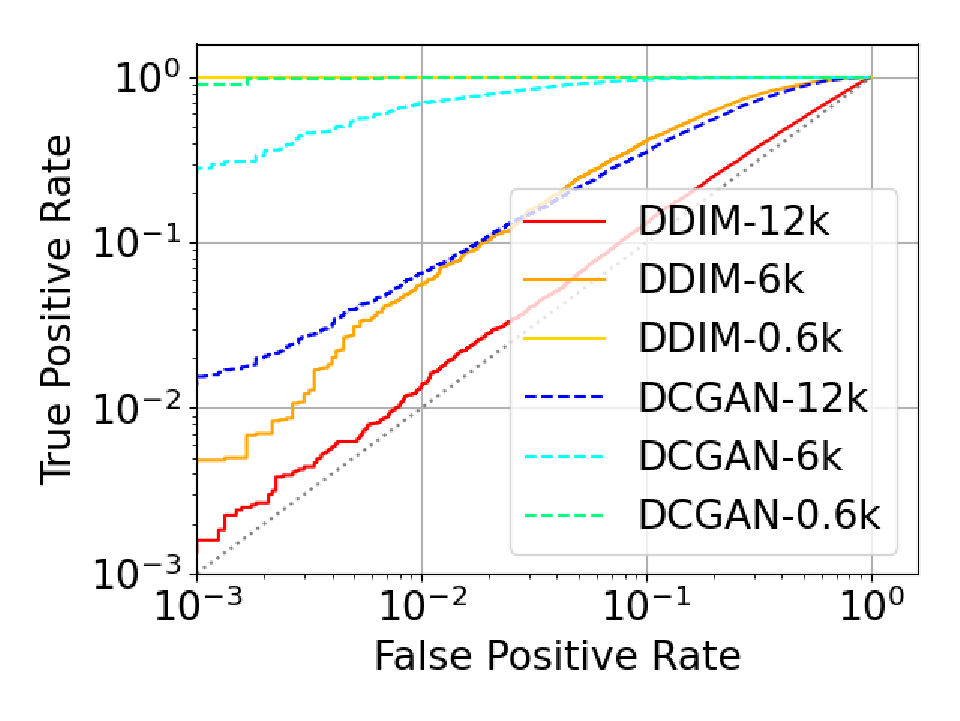}
        \caption{{\footnotesize White-box setting (CIFAR-10).}}
    \end{subfigure}
    \hfill
    \begin{subfigure}[b]{0.49\linewidth}  
        \centering 
        \includegraphics[keepaspectratio, scale=0.275]{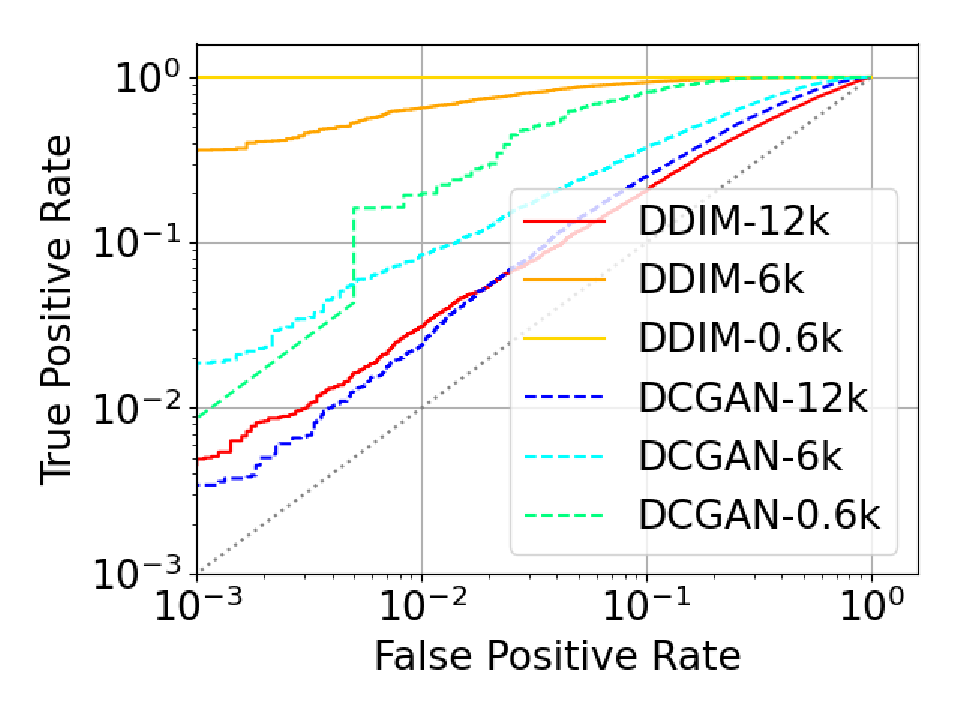}
        \caption{{\small White-box setting (CelebA).}}    
    \end{subfigure}
    \vskip\baselineskip
    \begin{subfigure}[b]{0.49\linewidth}   
        \centering 
        \includegraphics[keepaspectratio, scale=0.275]{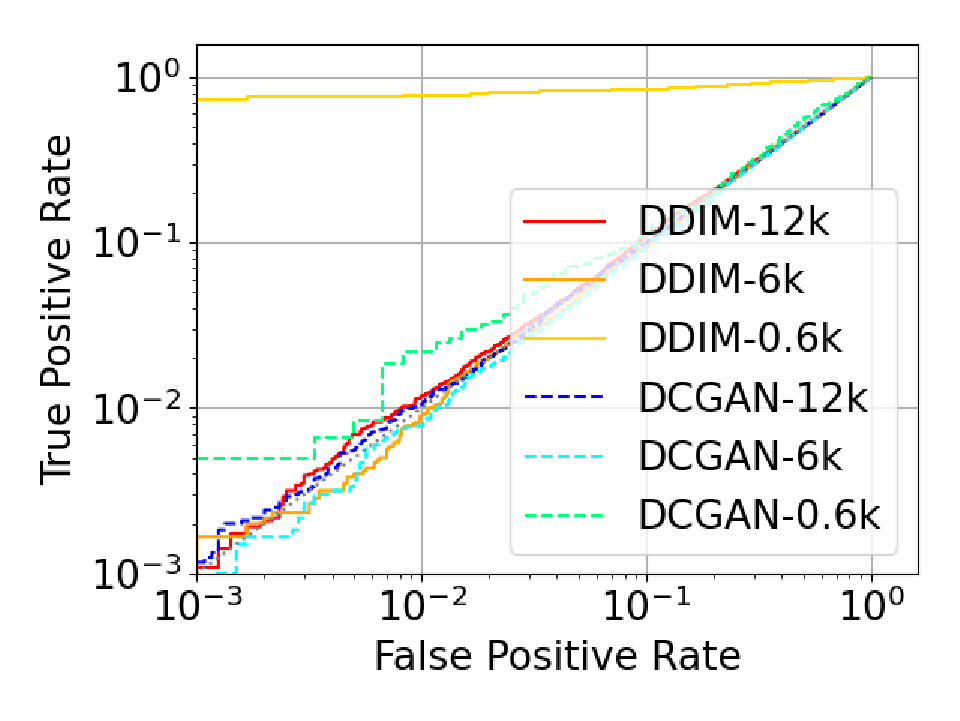}
        \caption{{\footnotesize Black-box setting (CIFAR-10).}}    
    \end{subfigure}
    \hfill
    \begin{subfigure}[b]{0.49\linewidth}   
        \centering 
        \includegraphics[keepaspectratio, scale=0.275]{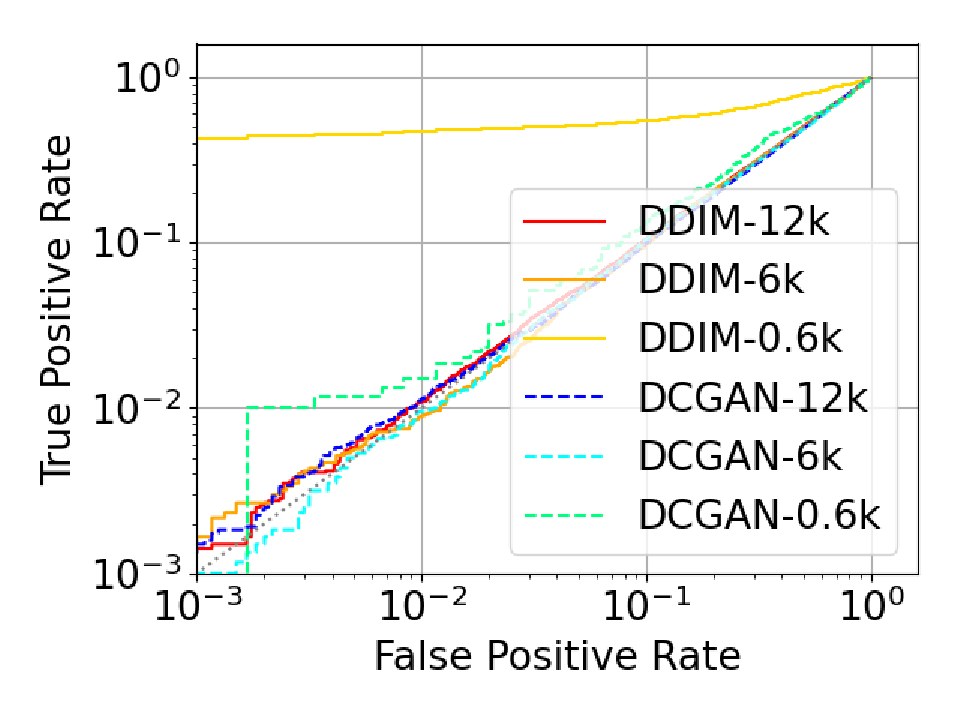}
        \caption{{\small Black-box setting (CelebA).}}    
    \end{subfigure}
    \caption{ROC curves for different number of training samples.} 
    \label{fig:roc_size}
\end{figure}

We perform the attacks on three target models, i.e., their number of training samples is 12,000, 6000, and 600 images, to evaluate the impact of number of training samples on a membership inference attack. 
The results are shown in Fig~\ref{fig:roc_size} and Table~\ref{tab:attack_eval_total}, where the number of epochs is changed depending on the number of training samples in order to align the number of images trained by the model. 
Overall, DDIM is vulnerable to membership inference attacks compared to DCGAN when the number of training samples is small. 
We analyze this result in detail below. 
We also note that FID for DCGAN on CIFAR-10 is high because DCGAN is unsuitable for the dataset. 
Indeed, we found many strange images, although we omit the detail due to the space limitation. 
Consequently, we focus on results on CelebA. 

First, in the white-box setting, Fig.~\ref{fig:roc_size} shows a remarkable phenomenon: 
DDIM is more vulnerable to the membership inference attack than DCGAN for small number of training samples but is more resistant for large number of training samples. 
For instance, DDIM is vulnerable in the white-box setting until 6000 images. 
After that, according to the columns of CelebA in Table~\ref{tab:attack_eval_total}, AUCROC and ASR of DDIM for 12,000 images are 0.064 smaller and 0.044 smaller than DCGAN, respectively.  

We also found remarkable results in the black-box setting: for small number of training samples, DDIM is significantly vulnerable compared to DCGAN. 
Notably, AUCROC and ASR of DDIM for 600 images are 0.24 higher and 0.19 higher than DCGAN, respectively. 
Meanwhile, after the number of training samples is 6,000, the result of DDIM for each model is identical to that of DCGAN for any metrics. 
It means that DDIM is vulnerable to membership inference attacks for small number of training samples instead of providing lower FID.

\subsubsection{Epochs}

\begin{figure}[t!]
  \begin{minipage}[b]{0.49\linewidth}
    \centering
    \includegraphics[keepaspectratio, scale=0.275]{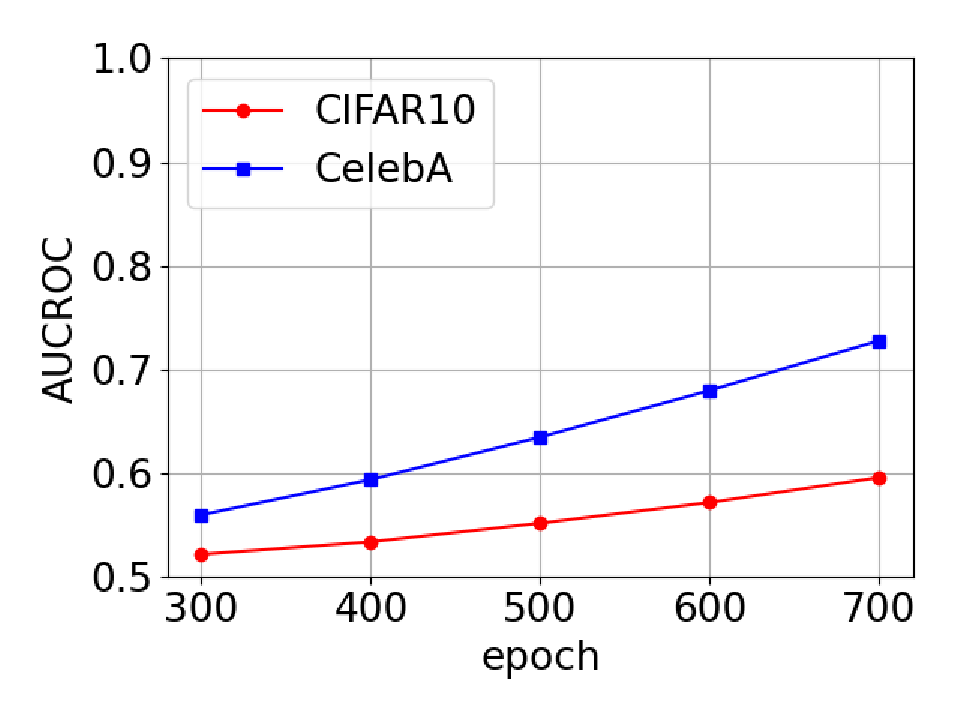}
    \subcaption{DDIM.}
  \end{minipage}
  \begin{minipage}[b]{0.49\linewidth}
    \centering
    \includegraphics[keepaspectratio, scale=0.275]{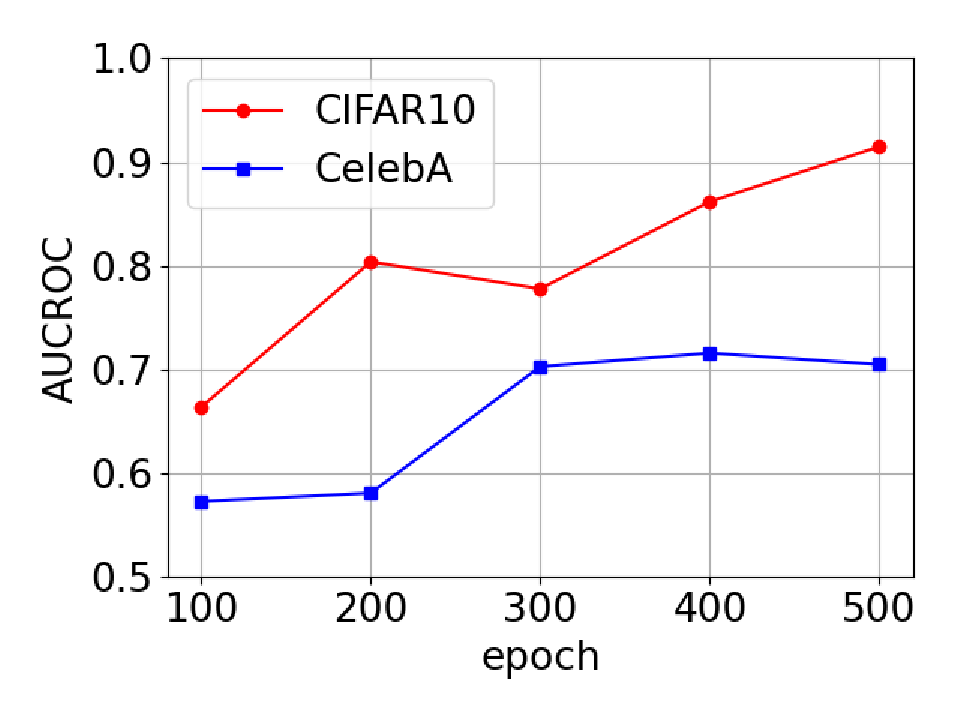}
    \subcaption{DCGAN.}
  \end{minipage}
  \caption{AUCROC throughout training on white-box setting.}
  \label{fig:auc_training}
\end{figure}

We perform the attacks on five target models, i.e., models with 300 epochs to 700 epochs, to evaluate the impact of the number of epochs  on a membership inference attack in the white-box setting. 
The results are shown in Fig~\ref{fig:auc_training}. 
We also note that we omit results in the black-box setting since those for DDIM and DCGAN are identical. 
According to the figure, values of AUCROC become higher in proportion to the number of epochs, and it is especially stable for DDIM. 
Since AUCROC of DDIM increases linearly, we can also estimate the number of epochs to obtain the resistance to a membership inference attack. 

\subsection{Impact of Hyperparameters for Sampling}

Hyperparameters often affect the model's performance, and hence we discuss the impact of hyperparameters for sampling, which are unique to diffusion models, i.e., sampling steps and sampling variances. 
Here, the investigation method in Section~\ref{sec:our_method} is independent of hyperparameters for sampling since they are used only for the image generation process. 
Consequently, we evaluate only in the black-box setting. 

\subsubsection{Sampling Steps}

\begin{figure}[t!]
  \begin{minipage}[b]{0.49\linewidth}
    \centering
    \includegraphics[keepaspectratio, scale=0.275]{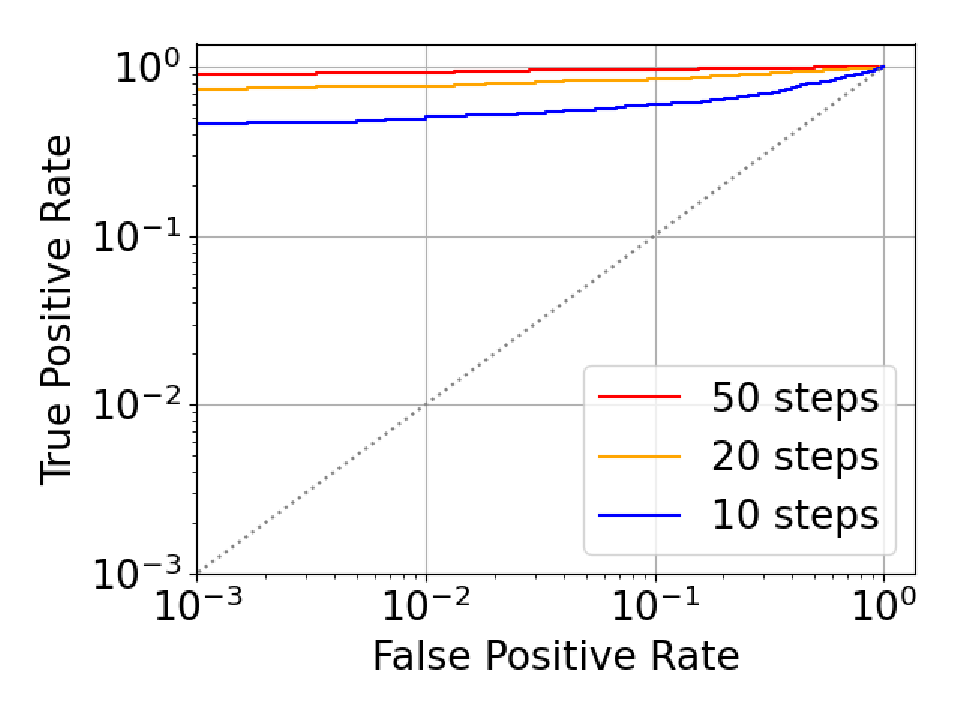}
    \subcaption{CIFAR-10.}
  \end{minipage}
  \begin{minipage}[b]{0.49\linewidth}
    \centering
    \includegraphics[keepaspectratio, scale=0.275]{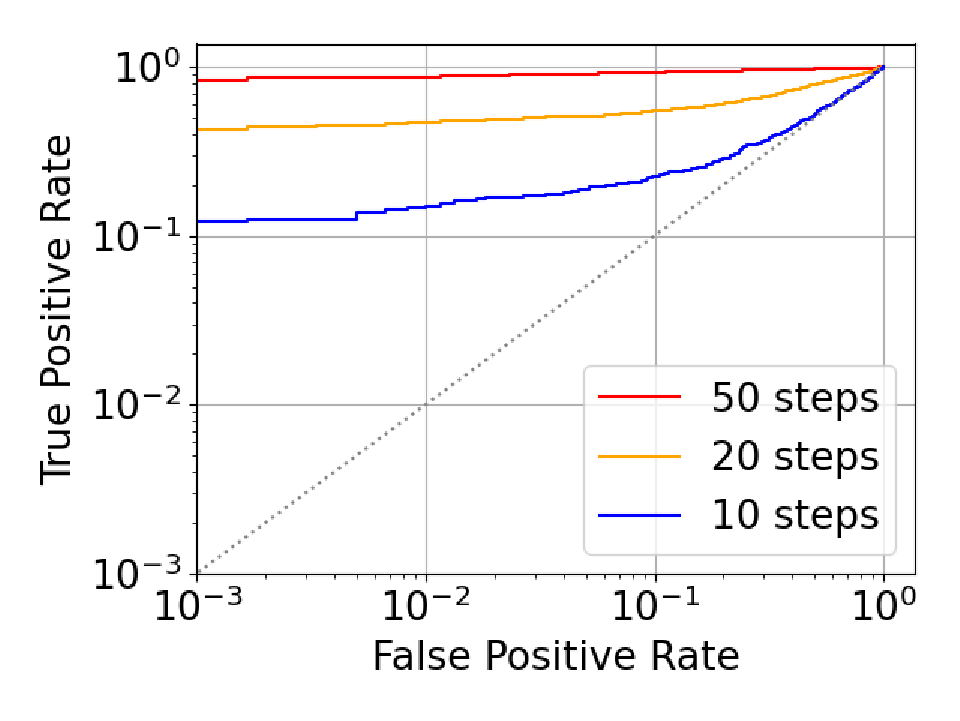}
    \subcaption{CelebA.}
  \end{minipage}
  \caption{ROC curves for different sampling steps on black-box setting.}
  \label{fig:roc_sampling}
\end{figure}

We perform the attacks on three different numbers of steps, i.e., 50 steps, 20 steps, and 10 steps, to evaluate the impact of sampling steps on a membership inference attack. 
The results are shown in Fig.~\ref{fig:roc_sampling} and Table~\ref{tab:sampling_step}, where the number of epochs is 1000 and the model is trained with 600 images. 
According to the figure, DDIM is more vulnerable to the membership inference attack in proportion to the number of sampling steps. 
The reason is that generated images have passed through the trained networks more frequently in proportion to the number of sampling steps. 
Namely, the generated images represent the training results strongly. 
It also means that sampling steps significantly affect the resistance to a membership inference attack. 

\begin{table}[t!]
    \centering
    \caption{Attack performance for different sampling steps on black-box setting.}
    \begin{tabular}{c|r|ccc|r}
    \hline
         Dataset & Steps & AUCROC & ASR & TPR at 1\%FPR & FID \\ \hline \hline
                  & 50 & 0.986 & 0.962 & 92.83\% & 22.60 \\
         CIFAR-10 & 20 & 0.931 & 0.882 & 77.50\% & 28.72 \\
                  & 10 & 0.781 & 0.724 & 51.17\% & 45.50 \\ \hline
         & 50 & 0.966 & 0.933 & 87.67\% & 12.37 \\
        CelebA & 20 & 0.762 & 0.717 & 47.33\% & 33.89 \\
                & 10 & 0.539 & 0.516 & 15.00\% & 101.01 \\ \hline
    \end{tabular}
    \label{tab:sampling_step}
\end{table}

\subsubsection{Sampling Variances} 

We perform the attacks on three different sampling variances, i.e., $\sigma(1)$, $\sigma(0)$, and $\hat{\sigma}$, to evaluate the impact of sampling variances that control the randomness of the sampling on a membership inference attack. 
The results are shown in Table~\ref{tab:sampling_var}, where the number of sampling steps is 20 and the model is trained with 600 images and 10,000 epochs. 
According to the table, although FID becomes larger in proportion to variances, DDIM is vulnerable to the membership inference attack for all variances. 
It means that the resistance to a membership inference attack is independent of sampling variances and they affect only the quality of generated images. 


\begin{table}[t!]
    \centering
    \caption{Attack performance for different sampling variances on black-box setting.}
    \begin{tabular}{c|c|ccc|r}
    \hline
         Dataset & Variance & AUCROC & ASR & TPR at 1\%FPR & FID \\ \hline \hline
                  & $\sigma(0)$ & 0.931 & 0.882 & 77.50\% & 28.72 \\
         CIFAR-10 & $\sigma(1)$ & 0.927 & 0.892 & 78.50\% & 39.10 \\
                  & $\hat{\sigma}$ & 0.929 & 0.886 & 79.83\% & 89.95 \\ \hline
                & $\sigma(0)$ & 0.762 & 0.717 & 47.33\% & 33.89 \\
         CelebA & $\sigma(1)$ & 0.799 & 0.763 & 55.50\% & 31.42 \\
                & $\hat{\sigma}$ & 0.803 & 0.759 & 54.17\% & 78.85 \\ \hline
    \end{tabular}
    \label{tab:sampling_var}
\end{table}

\subsubsection{Summary}

In summary, sampling steps are important for the resistance to a membership inference attack, while sampling variances are irrelevant. 
We leave it as an open problem to confirm if the same results are obtained in other architectures of diffusion models. 

\section{Conclusion}

In this paper, we investigated a membership inference attack on diffusion models. 
We primarily discussed the comparison of DDIM as a diffusion model with DCGAN as an existing generative model and hyperparameters of a diffusion model, i.e., timesteps, sampling steps, and sampling variances. 
We first showed that the diffusion model is comparably resistant to a membership inference attack as GAN through experiments on CelebA and CIFAR-10 datasets in the white-box and black-box settings. 
We then demonstrated the impact of timesteps on the membership inference attack. 
According to our result, intermediate steps in a noise schedule are the most vulnerable. 

We also found two key insights into a membership inference attack on the diffusion model through analysis for overfitting, including comparison with a GAN, and hyperparameters for sampling. 
First, for overfitting, we confirmed if DDIM is vulnerable to the attack when the sample sizes are small because it could be more well-trained than DCGAN. 
Second, sampling steps are significantly important for resistance to the attack, while sampling variances are irrelevant. 
We are in the process of investigating further resistance to membership inference attacks on diffusion models, including other architectures. 

\textbf{Acnowledgement:} 
A part of this work was supported by JST, CREST Grant JPMJCR21M5, Japan.

\bibliographystyle{IEEEtran}
\bibliography{main}


\end{document}